# Multi-View Graph Convolutional Network for Multimedia Recommendation

Penghang Yu
Nanjing University of Posts and Telecommunications
Nanjing, China
2022010201@njupt.edu.cn

Zhiyi Tan
Nanjing University of Posts and Telecommunications
Nanjing, China
tzy@njupt.edu.cn

Guanming Lu
Nanjing University of Posts and Telecommunications
Nanjing, China
lugm@njupt.edu.cn

Bing-Kun Bao*
Nanjing University of Posts and Telecommunications
Nanjing, China
Peng Cheng Laboratory
Shenzhen, China
bingkunbao@njupt.edu.cn

## ABSTRACT

Multimedia recommendation has received much attention in recent years. It models user preferences based on both behavior information and item multimodal information. Though current GCN-based methods achieve notable success, they suffer from two limitations: (1) **Modality noise contamination to the item representations.** Existing methods often mix modality features and behavior features in a single view (e.g., user-item view) for propagation, the noise in the modality features may be amplified and coupled with behavior features. In the end, it leads to poor feature discriminability; (2) **Incomplete user preference modeling caused by equal treatment of modality features.** Users often exhibit distinct modality preferences when purchasing different items. Equally fusing each modality feature ignores the relative importance among different modalities, leading to the suboptimal user preference modeling.

To tackle the above issues, we propose a novel **M**ulti-View **G**raph **C**onvolutional **N**etwork (**MGCN**) for the multimedia recommendation. Specifically, to avoid modality noise contamination, the modality features are first purified with the aid of item behavior information. Then, the purified modality features of items and behavior features are enriched in separate views, including the user-item view and the item-item view. In this way, the distinguishability of features is enhanced. Meanwhile, a behavior-aware fuser is designed to comprehensively model user preferences by adaptively learning the relative importance of different modality features. Furthermore, we equip the fuser with a self-supervised auxiliary task. This task is expected to maximize the mutual information between the fused multimodal features and behavior features, so as to capture complementary and supplementary preference information simultaneously. Extensive experiments on three public datasets demonstrate the effectiveness of our methods. Our code is made publicly available on https://github.com/demonph10/MGCN.

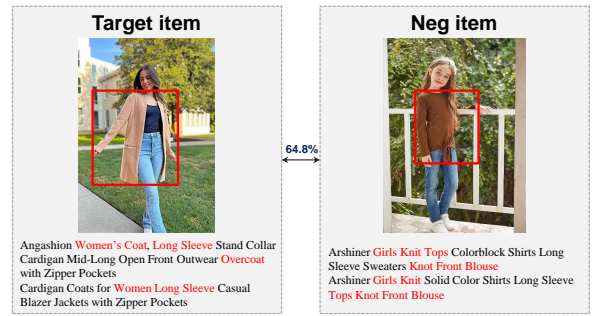

**Figure 1: Illustration of the modality noise contamination. The left image shows the product that user wants to purchase, while the right image is not. Red boxes/words illustrate the part that may be focused on. Due to contamination caused by the image background, the brightness, etc, the cosine similarity [15] of these two clothes reaches 64.8%.**

*Corresponding authors

## 1 INTRODUCTION

In contemporary times, recommender systems have attained widespread adoption in diverse domains [7, 23], with the objective of aiding users in discovering information that aligns with their preferences. As users' preferences are often influenced by multimodal information, researchers have begun to incorporate multimodal information into recommendation frameworks [41, 51]. A typical pipeline of multimedia recommendation is first extracting multimodal features by pre-trained deep neural networks, then fusing multimodal features with behavior features as the representations of items [2, 18]. Based on the multimodal representations of items, the user preferences are modeled through a carefully designed collaborative filtering recommendation framework. Since the collaborative signal is enhanced by multimodal features, the recommendation performance is thereby improved.

Early efforts in the multimedia recommendation, such as Visual Bayesian Personalized Ranking (VBPR) [10] and Cross-modal Knowledge Embedding (CKE) [47] extend Matrix Factorization (MF) [14] methods by incorporating visual features. As user interaction data can be naturally represented as a bipartite graph, recent researchers prefer to Graph Convolution Network (GCN) to capture high-order connectivity and enhance the preferences features [17, 38, 39, 43]. For example, MMGCN [39] incorporates different modality information into multiple user-item views, and models user preferences by concatenating the learned modality representation. Based on MMGCN, GRCN [38] utilizes multimodal features to refine user-item view, with the aim of pruning false-positive interaction. These GCN-based methods achieve great success and obtain state-of-the-art performance.

Despite the notable success, existing multimedia recommendation methods still suffer from two limitations: (1) **Modality noise contamination to the item representations.** There is plenty of preference-irrelevant modality noise contained in multimodal information [49, 51], such as the redundant text description, the image background, and the image brightness (e.g., which can be clearly seen in Figure.1). Directly injecting modality features would contaminate the representation learning of items. Even worse, current GCN-based [4, 21, 39] methods tend to mix modality features and behavior features in the user-item view for propagation. It means the noise would propagate between nodes, making the modality noise contamination further amplified and coupled with behavior features. In the end, the distinguishability of all node representations decreases. (2) **Incomplete user preference modeling caused by equal treatment of modality features.** Existing studies typically fuse the modality features by simple linear combination or concatenation, which treats each modality features equally [18, 35, 39]. Such way of modeling ignores the fact that users have different modality preferences when purchasing different items. In other words, users may pay attention to the thumbnail (i.e., visual information) when watching micro-video, while focusing on the description (i.e., textual information) when buying books. Current fusion mechanism cannot capture the relative importance of different modality features, thereby falling short in comprehensive modeling user preferences. These limitations lead to a suboptimal recommendation performance.

To tackle the above issues, we propose a novel **M**ulti-**V**iew **G**raph **C**onvolutional **N**etwork (**MGCN**) for the multimedia recommendation. The proposed model equips three specially designed modules: the Behavior-Guided Purifier, the Multi-View Information Encoder, and the Behavior-Aware Fuser. On one hand, to purify the modality information, a behavior-guided purifier is employed to denoise the modality features. It filters out preference-irrelevant features from the raw modality information, with the guidance of behavior information. Subsequently, the purified modality features are enriched by encoding semantically correlative signals under the item-item view, while user and item behavior features are enhanced by encoding the high-order collaborative signals under the user-item view. On the other hand, a behavior-aware fuser is developed to comprehensively model user preferences. It adaptively fuses items' modality features according to users' modality preferences, which are distilled from behavior features. A self-supervised auxiliary task is subsequently introduced during the fusion phase. The task is expected to maximize the mutual information between the fused multimodal features and behavior features, with the aim of simultaneously capturing complementary and supplementary features from both multimodal and behavior information. Comprehensive experiments on three public datasets demonstrate the distinct advantages of our methods.

Our main contributions can be summarized as follows:

- We develop a behavior-guided purifier, which effectively avoids the noise contamination issue with the guidance of behavior information.
- We design a multi-view information encoder, which enriches the representations by separately capturing high-order collaborative signals and semantically correlative signals.
- We propose a behavior-aware fuser and construct a novel self-supervised auxiliary task, which comprehensively models user preferences through adaptively fusing behavior information and multimodal information.

## 2 MODEL

### 2.1 Problem Definition

Let $\mathcal{U} = \{u\}$ donate the user set and $\mathcal{I} = \{i\}$ donate the item set. The input ID embeddings of user $u$ and item $i$ are $\mathbf{E}_{id} \in \mathbb{R}^{d \times (|U|+|I|)}$. $d$ is the embedding dimension. Then, we denote each item modality features as $\mathbf{E}_{i,m} \in \mathbb{R}^{d_m \times |I|}$, where $d_m$ is the dimension of the features, $m \in \mathcal{M}$ is the modality, and $\mathcal{M}$ is the set of modalities. In this paper, we mainly consider visual and textual modalities denoted by $\mathcal{M} = \{v, t\}$. Please kindly note that our method is not fixed to the two modalities and multiple modalities can be involved.

Next, user historical behavior data is denoted as $\mathcal{R} \in \{0, 1\}^{|U| \times |I|}$, where each entry $\mathcal{R}_{u,i} = 1$ if user $u$ clicked item $i$, otherwise $\mathcal{R}_{u,i} = 0$. Naturally, the historical interaction data $\mathcal{R}$ can be regarded as a sparse behavior graph $\mathcal{G} = \{\mathcal{V}, \mathcal{E}\}$, where $\mathcal{V} = \{\mathcal{U} \cup \mathcal{I}\}$ denotes the set of nodes and $\mathcal{E} = \{(u, i) | u \in \mathcal{U}, i \in \mathcal{I}, \mathcal{R}_{ui} = 1\}$ denotes the set of edges. The purpose of the multimedia recommendation is to accurately predict users' preferences by ranking items for each user according to predicted preferences scores $\hat{y}_{ui}$.

### 2.2 Behavior-Guided Purifier

Modality information provides rich and meaningful content information of items, while inevitably containing modality noise as well. To avoid noise contamination, we propose a behavior-guided purifier. Specifically, we first transform raw item modality features

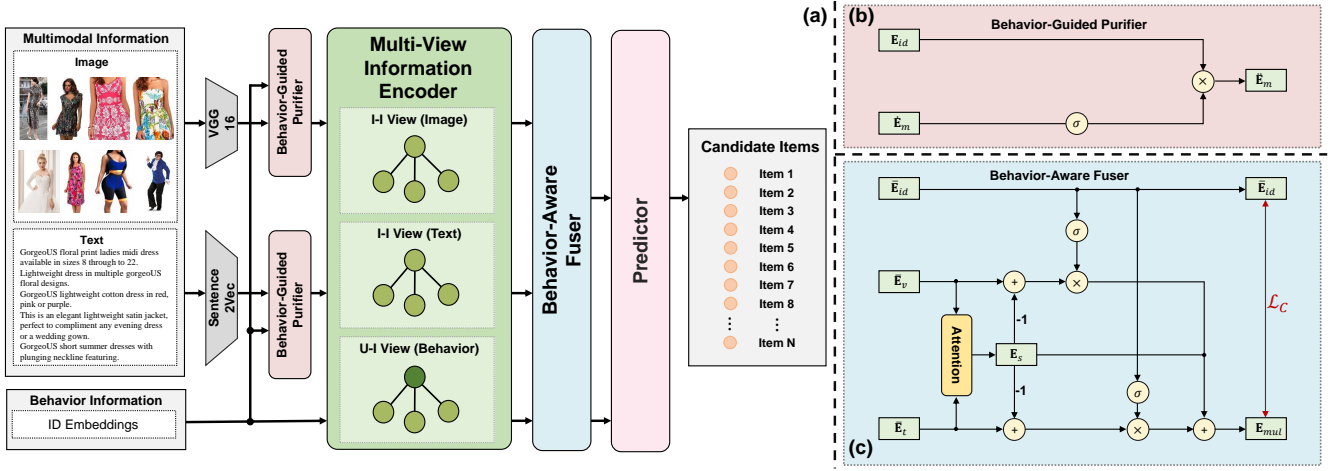

Figure 2: (a) The overall framework. (b) The behavior-guided purifier filters out modality noise with the help of behavior information. (c) The behavior-aware fuser adaptively fuses modality features according to user modality preferences.

$\mathbf{E}_{i,m}$ into high-level features $\mathbf{\dot{E}}_{i,m}$ :

$$\mathbf{\dot{E}}_{i,m} = \mathbf{W}_1 \mathbf{E}_{i,m} + \mathbf{b}_1, \tag{1}$$

where $\mathbf{W}_1 \in \mathbb{R}^{d \times d_m}$ and $\mathbf{b}_1 \in \mathbb{R}^d$ denote the trainable transformation matrix and the bias vector.

Then the preference-relevant modality features are separated from the modality features, with the guidance of behavior features:

$$\mathbf{\ddot{E}}_{i,m} = f_{gate}^m(\mathbf{E}_{i,id}, \mathbf{\dot{E}}_{i,m}) = \mathbf{E}_{i,id} \odot \sigma(\mathbf{W}_2 \mathbf{\dot{E}}_{i,m} + \mathbf{b}_2), \tag{2}$$

where $\mathbf{W}_2 \in \mathbb{R}^{d \times d}$ and $\mathbf{b}_2 \in \mathbb{R}^d$ are learnable parameters, $\odot$ represents the element-wise product and $\sigma$ is the sigmoid nonlinearity. With the guidance of behavior features that are encoded in the ID embedding $\mathbf{E}_{i,id}$, we separate the preference-relevant modality features $\mathbf{\ddot{E}}_{i,m}$ from item's representation $\mathbf{\dot{E}}_{i,m}$. For users, we obtain their modality features by aggregating the interacted items' modality features. We detail the process in Section 2.3.

## 2.3 Multi-View Information Encoder

According to [20, 30], both the collaborative signals and the semantically correlative signals can significantly influence the efficacy of multimedia recommendation. Thus, inspired by [30], we design a multi-view information encoder to enhance the discriminability of features. It captures collaborative signals from the view of the user-item relationship, and semantically correlative signals from the view of the item-item relationship.

### 2.3.1 User-Item View.
In particular, to capture high-order collaborative signals, we construct a GCN module to propagate ID embeddings of users and items over the interaction graph. The message propagation stage at $l$-th graph convolution layer can be formulated as:

$$\mathbf{E}_{id}^{(l)} = \mathbf{E}_{id}^{(l-1)} \mathcal{L}, \tag{3}$$

where $\mathbf{E}_{id}^{(l)}$ represents the enhanced representations of users and items in $l$-th graph convolution layer, $\mathbf{E}_{id}^{(0)}$ are the initial ID embeddings, that is $\mathbf{e}_u^{id\,(0)} = \mathbf{e}_u^{id}$ and $\mathbf{e}_i^{id\,(0)} = \mathbf{e}_i^{id}$. $\mathcal{L}$ represents the

Laplacian matrix for the user-item graph, which is formulated as:

$$\mathcal{L} = \mathbf{D}^{-\frac{1}{2}} \mathcal{A} \mathbf{D}^{-\frac{1}{2}}, \quad \text{and} \quad \mathcal{A} = \begin{vmatrix} 0 & \mathbf{R} \\ \mathbf{R}^{\mathsf{T}} & 0 \end{vmatrix}, \tag{4}$$

where $\mathcal{R}$ is the user-item interaction matrix, and $0$ is the all-zero matrix; $\mathcal{A}$ is the adjacency matrix and $\mathbf{D}$ is the diagonal degree matrix; as such, the nonzero off-diagonal entry $\mathcal{L}_{ui} = 1/\sqrt{|\mathcal{N}_u||\mathcal{N}_i|}$, which avoids the scale of embeddings increasing with graph convolution operations. $\mathcal{N}_u$ represents the set of user's $u$ neighbors, $\mathcal{N}_i$ represents the set of item's $u$ neighbors, $|\mathcal{N}_u|$ and $|\mathcal{N}_i|$ denote the size of $\mathcal{N}_u$ and $\mathcal{N}_i$.

The representations of the $l$-th layer encode the $l$-order neighbors' information. By aggregating high-order neighbor information, the final representations $\mathbf{\bar{E}}_{id}$ are obtained:

$$\mathbf{\bar{E}}_{id} = \frac{1}{L+1} \sum_{i=0}^{L} \mathbf{E}_{id}^{(l)}. \tag{5}$$

### 2.3.2 Item-Item View.
Similar to the view of user-item, graph convolution operations on item-item affinity graphs can capture semantically correlative signals, thereby enriching item modality features. However, propagating modality feature in a dense affinity graph is computationally demanding and may introduce noise through unimportant edges. Thus, we conduct KNN sparsification [5] on the dense graph.

Specifically, we first quantify the item-item affinities based on the similarity of each raw modality feature. Considering the computational complexity, cosine similarity has been selected. Then, a fully-connected graph is constructed, to indicate item-item affinities in modality $m$. The element in row $a$ and column $b$ of the affinity graph $\mathcal{S}_m$ is:

$$s_{a,b}^m = \frac{(e_a^m)^{\mathsf{T}} e_b^m}{\|e_a^m\| \|e_b^m\|}, \tag{6}$$

where $s_{a,b}^m$ represents the similarity between item $a$ and item $b$ in modality $m$.

By performing graph convolution operations, we capture the common modality features of adjacent nodes, which can be used to enhance the target node. To capture the most relevant features from neighbors, for each item $a$, we only preserve $K$ edges with the greatest similarity:

$$\dot{s}_{a,b}^m = \begin{cases} s_{a,b}^m, & s_{a,b}^m \in \text{top-}K(\{s_{a,c}, c \in \mathcal{I}\}), \\ 0, & \text{otherwise}, \end{cases} \quad (7)$$

where $\dot{s}_{a,b}^m$ represents the edge weight between item $a$ and item $b$ in $m$ modality. Same to the user-item view, we normalize the item-item affinity matrix to alleviate the exploding gradient problem:

$$\dot{S}_m = D_m^{-\frac{1}{2}} \dot{S}_m D_m^{-\frac{1}{2}}, \quad (8)$$

where $D^m$ is the diagonal degree matrix of $\dot{S}_m$. Then, we propagate all items' modality features $\tilde{E}_{i,m}$ through a GCN module on the corresponding item-item affinity matrix $\dot{S}_m$:

$$\tilde{E}_{i,m} = \dot{S}_m \tilde{E}_{i,m}. \quad (9)$$

It is capable of enriching the feature by capturing common features of similar items. However, it should be noted that in the item-item view, the semantic similarity of node modality features is significantly decreasing with the propagation path increasing. Stacking multiple graph convolution layers not only leads to the node over-smoothing issue, but also easily captures noisy features. Therefore, in this study, we constructed a shallow GCN module to propagate modality information over $\dot{S}_m$ (we set the graph convolution layer to 1, and prove the effect in Section 3.3).

Finally, we obtain user modality features by aggregating interacted item modality features. User $u$'s modality feature $\tilde{e}_{u,m}$ is expressed as:

$$\tilde{e}_{u,m} = \sum_{i \in \mathcal{N}_u} \frac{1}{\sqrt{|\mathcal{N}_u||\mathcal{N}_i|}} \tilde{e}_{i,m}. \quad (10)$$

By concatenating user modality features $\tilde{E}_{u,m}$ with item modality features $\tilde{E}_{i,m}$, the final modality feature $\tilde{E}_m \in \mathbb{R}^{d \times (|U|+|I|)}$ is obtained.

### 2.4 Behavior-Aware Fuser

To accurately capture items' features in different modalities, we design a behavior-aware fuser. It allows flexible fusion weight allocation based on user modality preferences, which can be distilled from the behavior features. Moreover, to encourage the model to comprehensively explore user preference, a self-supervised task is introduced in the fusion process. The task is expected to maximize the mutual information [12, 13] between behavior features and fused multimodal features.

Specifically, the modality preferences $P_m$ are first distilled from user behavior features:

$$P_m = \sigma(W_3 \tilde{E}_{id} + b_3), \quad (11)$$

where $W_3 \in \mathbb{R}^{d \times d}$ and $b_3 \in \mathbb{R}^d$ are learnable parameters, $\sigma$ is the sigmoid nonlinearity, which learns a nonlinear gate to model user modality features.

There are both modality-shared and modality-specific features possessed across all modalities. For modality-shared features, user attention remains consistent, as this aligns with the intended purpose of the user's purchase. For this reason, we first extract the modality-shared features through attention mechanisms [28, 35], where the attention weight of each modality features $\tilde{E}_m$ are calculated as:

$$\alpha_m = \text{softmax}(q_1^T \tanh(W_4 \tilde{E}_m + b_4)), \quad (12)$$

where $q_1 \in \mathbb{R}^d$ denotes attention vector and $W_4 \in \mathbb{R}^{d \times d}$, $b_4 \in \mathbb{R}^d$ denote the weight matrix and the bias vector, respectively. Notice that these parameters are shared for all modalities. Then, the modality-shared features $E_s$ are obtained:

$$E_s = \sum_{m \in \mathcal{M}} \alpha_m \tilde{E}_m. \quad (13)$$

Then, the modality-specific features $\tilde{E}_m$ are obtained by subtracting the modality-shared features $E_s$:

$$\tilde{E}_m = \tilde{E}_m - E_s. \quad (14)$$

Finally, we adaptively fuse the modality-specific features $\tilde{E}_m$, and combine them with modality-shared features $E_s$ as the final features $E_{mul}$:

$$E_{mul} = E_s + \frac{1}{|\mathcal{M}|} \sum_{m \in \mathcal{M}} \tilde{E}_m \odot P_m. \quad (15)$$

In order to promote the exploration of behavior and multimodal information, a self-supervised auxiliary task has been devised. The mathematical expression of this task is as follows:

$$\begin{aligned} \mathcal{L}_C = & \sum_{u \in \mathcal{U}} -\log \frac{\exp(e_{u,mul} \cdot \tilde{e}_{u,id}/\tau)}{\sum_{v \in \mathcal{U}} \exp(e_{v,mul} \cdot \tilde{e}_{v,id}/\tau)} \\ & + \sum_{i \in \mathcal{I}} -\log \frac{\exp(e_{i,mul} \cdot \tilde{e}_{i,id}/\tau)}{\sum_{j \in \mathcal{I}} \exp(e_{j,mul} \cdot \tilde{e}_{j,id}/\tau)}, \end{aligned} \quad (16)$$

where $\tau$ is the temperature hyper-parameter of softmax.

### 2.5 Predictor

Based on the enhanced behavior features and multimodal features, we form the final representations of users and items:

$$\begin{aligned} e_u &= \tilde{e}_{u,id} + e_{u,mul}, \\ e_i &= \tilde{e}_{i,id} + e_{i,mul}. \end{aligned} \quad (17)$$

Followed [11], the inner product is adopted to determine the likelihood of interaction between user $u$ and item $i$:

$$f_{predict}(u, i) = \hat{y}_{ui} = e_u^T e_i. \quad (18)$$

### 2.6 Optimization

During the phase of model training, we adopt the Bayesian Personalized Ranking (BPR) loss $\mathcal{L}_{BPR}$ as the basic optimization task, which assumes that users prefer historically interacted items over unclicked ones. And it is combined with auxiliary self-supervised tasks to jointly update the representations of users and items:

$$\mathcal{L} = \mathcal{L}_{BPR} + \lambda_C \mathcal{L}_C + \lambda_E \|E\|_2, \quad (19)$$

where $E$ is the set of model parameters; $\lambda_C$ and $\lambda_E$ are hyperparameters to control the effect of the contrastive auxiliary task and the $L_2$ regularization, respectively.

**Table 1: Statistics of the experimental datasets**

| Dataset | #User | #Item | #behavior | Density |
|---------|-------|-------|-----------|---------|
| Baby | 19,445 | 7,050 | 160,792 | 0.117% |
| Sports | 35,598 | 18,357 | 296,337 | 0.045% |
| Clothing | 39,387 | 23,033 | 278,677 | 0.031% |

## 3 EXPERIMENTS

In this section, we conduct extensive experiments to evaluate the performance of the proposed MGCN model on three public datasets. The following four questions can be well answered through experiment results:

- **RQ1**: How does MGCN perform compared with the state-of-the-art multimedia recommendation methods and other collaborative filtering methods?
- **RQ2**: How do the modules influence the performance of MGCN?
- **RQ3**: How do different hyper-parameter settings impact the results of the MGCN model?
- **RQ4**: Why purifying modality information can achieve better recommendation performance?

### 3.1 Experimental Settings

*3.1.1 Dataset.* Following [48, 53], we conduct experiments on three categories of the widely used Amazon dataset[1]: (a)Baby, (b) Sports and Outdoors, and (c) Clothing, Shoes, and Jewelry, which we refer to as Baby, Sports, and Clothing in brief. The statistics of these datasets are presented in Table.1. Following [53], we use the pre-extracted 4,096-dimensional visual features and 384-dimensional text features, which have been published in [52].

*3.1.2 Compared Methods.* To evaluate the effectiveness of our proposed model, we compare it with several representative recommendation models. These baselines fall into two groups: General models, which only rely on interactive data for recommendation; Multimedia models, which utilize both interactive data and multi-modal features for the recommendation.

**i) General Models:**

- **MF** [14]: This is a classic collaborative filtering method, which learns user and item representations with a matrix factorization framework.
- **LightGCN** [11]: This is the most popular GCN-based collaborative filtering method, which simplifies the design of GCN to make it more appropriate for the recommendation.

**ii) Multimedia Models:**

- **VBPR** [10]: This is a classic multimedia filtering method, which integrates the visual features and ID embeddings of each item as its representation. This can be seen as an extension of the MF model [14].
- **MMGCN** [39]: This method constructs a modal-specific graph to learn different modality features. It concatenates all modality features to obtain the representations of users or items for prediction.

- **GRCN** [38]: This method improves previous GCN-based models by refining the user-item interaction graph. With the help of multimodal features, the false-positive interaction can be identified and removed.
- **SLMRec** [24]: This method devises a self-supervised framework for multimedia recommendation. It constructs a node self-discrimination task, attempting to uncover the item multimodal pattern.
- **BM3** [53]: This method simplifies the self-supervised framework. It removes the requirement of randomly sampled negative examples and directly perturbs the representation through a dropout mechanism.
- **MICRO** [49]: This method is an extension of the state-of-the-art method LATTICE [48], which mines the latent structure between items by learning an item-item graph from their multimodal features.

*3.1.3 Evaluation Protocols.* For a fair comparison, we follow the same evaluation setting of [49, 53] with a random data splitting 8:1:1 on the interaction history of each user for training, validation, and testing. Besides, we follow the all-ranking protocol to evaluate the top-K recommendation performance and report the average metrics for all users in the test set: Recall@$K$ and NDCG@$K$.

*3.1.4 Implementation Details.* We implement the proposed model and all the baselines with MMRec [2] [52], which is a unified open-source framework to develop and reproduce recommendation algorithms. To ensure a fair comparison, we optimize all the methods with Adam optimizer and referred to the best hyperparameter settings reported in the original baseline papers. As for the general settings, For the general settings, we initialized the embedding with Xavier initialization of dimension 64, set the regularization coefficient to $\lambda_E = 10^{-4}$, and the batch size set to $B = 2048$. For the self-supervised task, we set the temperature $\tau = 0.2$, which is commonly considered a great choice. For convergence consideration, the early stopping and total epochs are fixed at 20 and 1000, respectively. Following [49], we use Recall@20 on the validation data as the training-stopping indicator.

### 3.2 Overall Performance (RQ1)

Table.2 shows the performance comparison of the proposed MGCN and other baseline methods on three datasets. From the table, we find several observations:

(1) **MGCN significantly outperforms both general recommendation models and multimedia recommendation models.** This indicates that our proposed method is well-designed for the multimedia recommendation. Specifically, instead of directly incorporating modality information, we first purified it with the guidance of behavior information. This avoids contamination from modality noise. Besides, through the multi-view information encoder, the behavior features and modality features are enriched, by encoding the high-order collaborative signal and semantically correlative signals. We then obtain each user's and item's representations through a behavior-aware fuser, which adaptively fuses modality features according to user modality preference. Moreover, we introduce a self-supervised task to maximize the mutual information between behavior features and fused multimodal features. This allows the model effectively extract user preferences from the various features, rather than seeking shortcuts from behavior or multimodal features. As a result, MGCN outperforms existing methods on all three datasets. Especially on the Clothing dataset, MGCN achieves 23.3% improvement over the best baseline methods.

**Table 2: Performance Comparison of Different Recommendation Models**

| Datasets | Metrics | MF<br>UAI'09 | LightGCN<br>SIGIR'20 | VBPR<br>AAAI'16 | MMGCN<br>MM'19 | GRCN<br>MM'20 | SLMRec<br>TMM'22 | BM3<br>WWW'23 | MICRO<br>TKDE'22 | Ours |
|---|---|---|---|---|---|---|---|---|---|---|
| Baby | Recall@10 | 0.0357 | 0.0479 | 0.0423 | 0.0378 | 0.0532 | 0.0540 | 0.0564 | <u>0.0584</u> | **0.0620** |
| | Recall@20 | 0.0575 | 0.0754 | 0.0663 | 0.0615 | 0.0824 | 0.0810 | 0.0883 | <u>0.0929</u> | **0.0964** |
| | NDCG@10 | 0.0192 | 0.0257 | 0.0223 | 0.0200 | 0.0282 | 0.0285 | 0.0301 | <u>0.0318</u> | **0.0339** |
| | NDCG@20 | 0.0249 | 0.0328 | 0.0284 | 0.0261 | 0.0358 | 0.0357 | 0.0383 | <u>0.0407</u> | **0.0427** |
| Sports | Recall@10 | 0.0432 | 0.0569 | 0.0558 | 0.0370 | 0.0559 | 0.0676 | 0.0656 | <u>0.0679</u> | **0.0729** |
| | Recall@20 | 0.0653 | 0.0864 | 0.0856 | 0.0605 | 0.0877 | 0.1017 | 0.0980 | <u>0.1050</u> | **0.1106** |
| | NDCG@10 | 0.0241 | 0.0311 | 0.0307 | 0.0193 | 0.0306 | 0.0374 | 0.0355 | <u>0.0367</u> | **0.0397** |
| | NDCG@20 | 0.0298 | 0.0387 | 0.0384 | 0.0254 | 0.0389 | 0.0462 | 0.0438 | <u>0.0463</u> | **0.0496** |
| Clothing | Recall@10 | 0.0187 | 0.0340 | 0.0280 | 0.0197 | 0.0424 | 0.0452 | 0.0421 | <u>0.0521</u> | **0.0641** |
| | Recall@20 | 0.0279 | 0.0526 | 0.0414 | 0.0328 | 0.0650 | 0.0675 | 0.0625 | <u>0.0772</u> | **0.0945** |
| | NDCG@10 | 0.0103 | 0.0188 | 0.0159 | 0.0101 | 0.0225 | 0.0247 | 0.0228 | <u>0.0283</u> | **0.0347** |
| | NDCG@20 | 0.0126 | 0.0236 | 0.0193 | 0.0135 | 0.0283 | 0.0303 | 0.0280 | <u>0.0347</u> | **0.0428** |

(2) **Compared to the MF-based methods, GCN-based methods are more vulnerable to modality noise contamination.** Specifically, VBPR outperforms MF by directly concatenating modal and behavioral features, while MMGCN performs worse than Light-GCN. This is mainly because of the message propagation mechanism in GCN-based methods, which continuously spreads noise and contaminates the representations of all users and items. Although some self-supervised methods (such as SLMRec and BM3) have constructed a self-discriminating task for nodes and partially alleviated the problem of modality noise contamination by eliminating redundant modality information. Nonetheless, these methods are still prone to modality noise contamination. In contrast, our proposed method employs a behavior-guided purifier to fundamentally filter out modality noise. As a result, our method has achieved the best recommendation performance.

(3) **An indirect injection of modality features may mitigate the issue of modality noise contamination.** Different from injecting modality features into representations, GRCN exclusively utilizes modality features to refine the user-item interaction graph. By uncovering latent item semantic relationships, MICRO constructed auxiliary item-item graphs for message propagation. However, both GRCN and MICRO have some limitations. First, they require frequent graph structure updates during training, which leads to high computational complexity and memory consumption. Additionally, modality information is not incorporated into representation learning, which limits the ability to comprehensively model user preferences. In contrast, our proposed method tackles the noise contamination issue via a purifier. The purified modality features are then enriched under different views. This allows for fine-grained fusion with behavior features, ultimately leading to the best results.

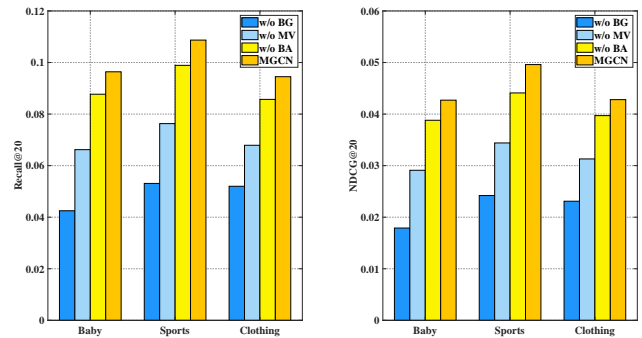

**Figure 3: Performance Comparison between different variants of MGCN. (Blue: w/o BG, Light Blue: w/o MV, Yellow: w/o BA, Orange: MGCN)**

## 3.3 Ablation Study (RQ2)

To comprehensively investigate the effects of various factors, we perform ablation studies on both the modules of MGCN and modality features.

### 3.3.1 Effect of Modules.
To investigate the effects of the keys components in MGCN, we set up the following model variants:

- **w/o BG**: We remove the behavior-guided purifier. Instead, the pre-extracted modality features are directly fed into the multi-view information encoder.
- **w/o MV**: We remove the multi-view information encoder. The modality features are concatenated and propagated on the user-item interaction graph. It is equally to simply encode collaborative signals.
- **w/o BA**: We remove the behavior-aware fuser. The final representations are obtained by averaging each modality feature $\tilde{E}_m$ and behavior features $\tilde{E}_{id}$.

Figure. 3 records Recall@20 of these variants on three datasets, and we have the following finds:

**Table 3: Performance Comparison under different modalities**

| Datasets | Modality | Recall@20 | NDCG@20 |
|----------|----------|-----------|---------|
| **Baby** | Text | 0.0857 | 0.0386 |
| | Visual | 0.0939 | 0.0408 |
| | All | **0.0964** | **0.0427** |
| **Sports** | Text | 0.1017 | 0.0454 |
| | Visual | 0.1055 | 0.0465 |
| | All | **0.1106** | **0.0496** |
| **Clothing** | Text | 0.0902 | 0.0404 |
| | Visual | 0.0919 | 0.0422 |
| | All | **0.0945** | **0.0428** |

(1) On the three datasets, models w/o BG significantly underperforms MGCN. It indicates that directly incorporating pre-extracted modality features is not a suitable approach for recommendation tasks, as modality information contains a large amount of preference-irrelevant noise. Through the behavior-guided purifier, the modality noise is effectively filtered out. Only the preferences-relevant modality features would be retained. This avoids the issue of modality noise contamination.

(2) Semantic correlation signals also have a significant impact on recommendation performance, indicating the importance of multi-view information encoders. Previous studies have also shown the importance of exploiting item-item relationships, which can effectively alleviate the data sparsity issue.

(3) The model with a behavior-aware fuser achieves better recommendation performance. That's mainly because users have different modality preferences when buying different items. Although behavior-unknown fusion mechanisms such as feature concatenation can also achieve modeling of user preferences, they overlook the fact that the importance of different modalities of items is different, which is not sufficient to obtain powerful representations. Thereby leading to a suboptimal recommendation performance.

*3.3.2 Effect of Modalities.* To explore the contribution of each modality to the recommendation performance, we conduct experiments under different input conditions. *Text* includes text information and behavior information, *Visual* includes visual information and behavior information, *All* includes all modality information and behavior information. As shown in Table. 3, it indicates that both textual and visual features can improve performance, but visual features have a greater impact. We attribute this to the fact that users tend to purchase items based on their appearance. Additionally, text descriptions often contain irrelevant information and can be overwhelming for users. To validate this hypothesis, we conducted a visualization analysis in Section 3.5.

## 3.4 Sensitivity Analysis (RQ3)

*3.4.1 Effects of the number of item neighbor $k$.* To avoid propagating the message from the irrelevant items, we constructed the item-item affinity graph with only the $k$ most similar items. Our experiments have shown that $k = 15$ is typically the most appropriate value for the number of item neighbor $k$. However, for the Baby dataset, $k = 20$ is more suitable. The optimal value of $k$ may vary

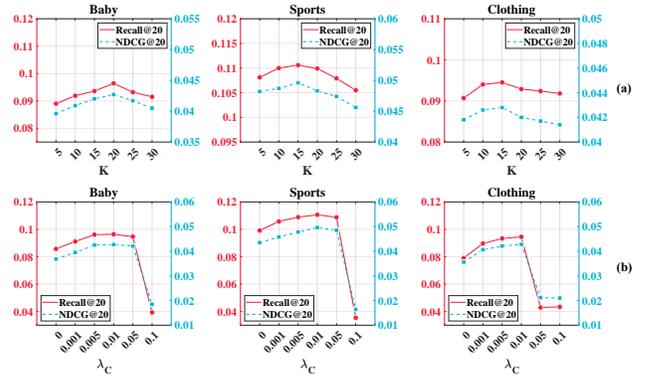

**Figure 4: Performance comparison w.r.t. different (a) numbers of neighbor $k$ and (b) weights of self-supervised task $\lambda_C$.**

depending on the scenario, but using a smaller value can reduce noise from unrelated neighbors.

*3.4.2 Effects of the weight of self-supervised task $\lambda_C$.* We investigated the impact of self-supervised auxiliary tasks and found consistent results across three datasets, as presented in Figure 4. The results show that jointly optimizing the primary recommendation task with a self-supervised auxiliary task leads to improvements in performance. We found that the optimal value of $\lambda_C$ is approximately 0.01, beyond which performance significantly declines. This suggests that a small $\lambda_C$ would promote the primary task. However, when $\lambda_C$ is too large, the model focuses excessively on the auxiliary task, and the model is misled by the self-supervised task. Therefore, finding a suitable auxiliary loss weight is crucial for ensuring the effectiveness of the model's recommendation.

## 3.5 Visualization Analysis (RQ4)

Intuitively, purifying modality information can resist modality noise contamination. To better understand the advantages of purifying modality information, we visualize the distribution of representations in *Clothing* Dataset. Figure. 5 and 6 demonstrates the impact of the behavior-guided purifier on representation learning. In particular, we randomly sample 500 items from the Clothing dataset and map their representations to a 2-dimensional space using t-SNE [27]. Next, we plot the 2D feature distributions using Gaussian kernel density estimation (KDE) [26]. For a clearer presentation, the Gaussian kernel density estimation of $arctan(y, x)$ on the unit hypersphere $\mathcal{S}^1$ is visualized in the bottom of each figure.

Upon analyzing the distribution of 2D features, we discovered that raw modality representations display multiple community clusters and an uneven single peak pattern in kernel density estimation. Directly fusing such low discriminative modality features with behavior features would result in low distinguishability, thereby decreasing the recommendation preference. The distribution of textual features is more uneven than that of visual features, which explains why using textual features is less effective than using visual features. By purifying modality features with the aid of behavior information, we avoid the inclusion of modality noise, resulting in enhanced discrimination of nodes' representations. Previous research [29] has demonstrated that representation uniformity is the key factor affecting recommendation performance. It explains why MGCN outperforms other multimedia recommendation methods.

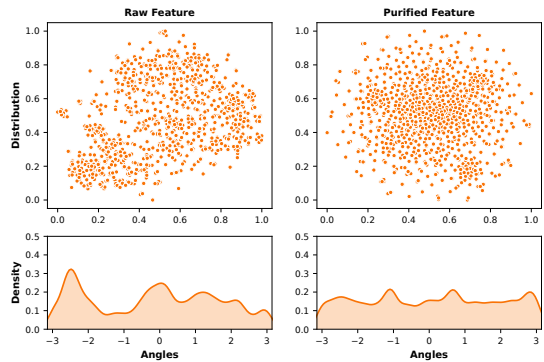

**Figure 5: The distribution of representations in text modality. The left of the figure shows the distribution of raw features, while the right displays the distribution of purified features.**

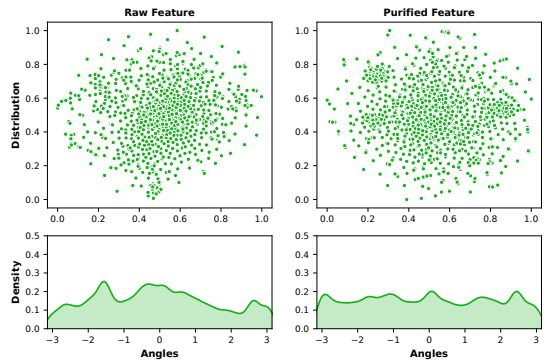

**Figure 6: The distribution of representations in visual modality. The left of the figure shows the distribution of raw features, while the right displays the distribution of purified features.**

## 4 RELATED WORK

### 4.1 Multimedia Recommendation

Collaborative Filtering (CF) has emerged as a prominent recommendation method that relies on behavior similarity to make top-k recommendations [19, 22]. However, users' preferences are often influenced by multimodal information, prompting researchers to incorporate it to improve CF-based approaches [41, 51]. Typically, multimodal features are extracted through pre-trained neural networks, which are then fused with behavior features to better model user preferences. For instance, VBPR [10] and VPOI [32] use convolution neural networks (CNN) pre-trained on ImageNet to extract deep visual features and enrich item representations. In addition to using only visual modality, some methods incorporate multimodal features into item representations. MDCF [44] maps multimodal features to a consensus Hamming space for the cold-start recommendation, while MV-RNN [8] uses multimodal features for sequential recommendation in a recurrent framework. However, these methods overlook the presence of preference-irrelevant features in modality information. Direct utilization of modality information may contaminate the item representations.

### 4.2 Graph Convolution Network

As user behavior data (e.g., clicks or purchases) can be naturally represented as a bipartite graph, recent researchers have favored Graph Convolution Network (GCN) [6, 20, 42] as a powerful tool to extract user behavior features. Specifically, NGCF [33] captures user behavior features by iteratively performing neighbor aggregation in the user-item view. LightGCN [11] simplifies the classical graph convolution module, making it more suitable for recommendation scenarios. Inspired by these works, GCN-based methods have also been applied to multimedia recommendation tasks [31, 36, 38, 39, 49]. For example, MMGCN [39] constructs multiple GCN modules to process different modalities, and concatenates the obtained modality features as the final representation of items. However, the message propagation mechanism in the GCN modules causes modality noise to propagate in the whole graph. Different from directly incorporating modality features, GRCN [38] and MICRO [49] adopt a graph structure learning module to learn potential semantic structures as a supplement to behavior information. However, not injecting modality information is not conducive to fully exploring user preferences. In addition, the graph structure learning module requires complex graph update operations, which can be costly for large-scale datasets. Moreover, existing methods overlook the fact that users have varying levels of attention to different modalities when purchasing different products, and equal treatment of each modality feature is not sufficient to fully explore user preferences.

## 5 CONCLUSION

In this paper, we have proposed a **M**ulti-View **G**raph **C**onvolutional **N**etwork (**MGCN**) for the multimedia recommendation. Specifically, we first develop a behavior-guided purifier to avoid modality noise contamination. Then the purified features and behavior features are separately enriched through a multi-view information encoder. Meanwhile, to comprehensively model user preferences, we design a behavior-aware fuser and propose a novel self-supervised auxiliary task.

In our future work, we aim to capture user preferences by integrating external knowledge with item information through large-scale language models. We posit that the development of such models could potentially address the cold-start problem.

## 6 ACKNOWLEDGMENT

This work was supported by National Key Research and Development Project (No.2020AAA0106200), the National Nature Science Foundation of China under Grants (No.61936005, No.62325206),the Natural Science Foundation of Jiangsu Province (No.BK20200037), and the Graduate research and innovation projects in Jiangsu Province (KYCX23_1026).